# Assessing the physical risks of climate change for the financial sector: a case study from Mexico's Central Bank

Francisco Estrada[1,2,3*], Miguel A. Altamirano del Carmen[3], Oscar Calderón-Bustamante[2], W.J. Wouter Botzen[2], Serafín Martínez-Jaramillo[4] & Stefano Battiston[5,6]

[1]*Instituto de Ciencias de la Atmósfera y Cambio Climático, Universidad Nacional Autónoma de México, Ciudad Universitaria, Circuito Exterior, 04510 Mexico, DF, Mexico,* [2]*Institute for Environmental Studies, Vrije Universiteit, Amsterdam, Netherlands,* [3]*Programa de Investigación en Cambio Climático, Universidad Nacional Autónoma de México, Ciudad Universitaria, Circuito Exterior, 04510 Mexico, DF, Mexico,* [4]*World Bank, USA,* [5]*University of Zurich, Zurich, Switzerland,* [6]*Ca' Foscari University of Venice, Italy.*

*Corresponding author, feporrua@atmosfera.unam.mx*

**Abstract**

The financial sector is increasingly concerned with the physical risks of climate change, but economic and financial impact representations are still developing, particularly for chronic risks. Mexico's Central Bank conducted a comprehensive assessment using a suite of global models to evaluate both physical and transition risks. We present the analysis concerning with chronic physical risks, underlining innovations such as the use of a recent integrated assessment model that enables grid-cell level analysis and differentiates urban and non-urban areas, capturing the local effects of climate change more accurately. The model includes multiple damage functions and a probabilistic climate model for encompassing analyses and detailed economic impact insights. Under the Current Policies scenario, economic losses could exceed 35% of Mexico's GDP by 2100. Accounting for the urban heat island effect, losses could surpass $20 trillion in present value, over ten times Mexico's 2024 GDP. However, implementing a scenario aligned with the Paris Agreement significantly reduces these losses, showcasing the benefits of international mitigation efforts, though substantial residual impacts persist. This study emphasizes integrating chronic physical risks into financial evaluations, proposing new approaches, metrics, and methods that exploit detailed, spatially explicit measures to improve risk and loss estimation and facilitate communication.

1. Introduction

There is an expanding interest in assessments of physical risks of climate change for financial sector organizations. This interest is motivated by increasing historical natural hazard losses around the world that already adversely affect the financial sector, notably the insurance industry (Hoeppe, 2016; Botzen, 2021). Moreover, future projections of climate change that involve an increasing frequency and/or severity of extreme weather events as well as gradual warming and sea level rise could inflict severe economic losses (van der Wijst et al., 2023), which in turn pose future risks for the financial sector. Relevant consequences for financial sector organizations include rising insured losses from natural hazards, decreasing stock and bond prices in financial markets, non-performing loans for banks, declining collateral values for investors, and worse macroeconomic conditions, amongst others (Zhou et al., 2023; Reinders et al., 2023). In addition to these physical risks, climate change can pose transition risks of moving towards a low-carbon economy, for example when investments in fossil-fuel based economic activities reduce in value (Monasterolo, 2020). These risk projections have spurred voluntary climate-related financial disclosures by financial sector regulations (TCFD, 2017), and steps towards increasingly stringent regulations of disclosures of climate risks for financial institutions, such as in the EU (European Central Bank, 2022).

An emerging literature reviewed by Zhou et al. (2023) and Reinders et al. (2023) has examined various climate related impacts to different components of the financial sector. Most of these studies have used econometric analysis to examine the influence of historical extreme weather conditions on financial sector indicators, such as bank stability (Blickle et al., 2021) and credit supply (Brei et al., 2019), and stock (Lanfear et al., 2019) and bond market returns (Huynh and Xia, 2021). An advantage of these statistical studies is that they offer empirical evidence on how climate related shock may influence financial sector performance. However, few of these studies provide a forward-looking perspective of how climate change affects future risks for the financial sector (Zhou et al., 2023). Such a forward-looking perspective is more commonly provided by computation models that assess future climate impacts under scenarios of climate and socio-economic change. These computational models are more often applied to assess climate change risks for the insurance industry (Zhou et al., 2023). Examples are natural catastrophe models that assess how more frequent and/or severe future natural hazard conditions affect insured losses (Kunreuther et al., 2013) or partial equilibrium models that assess how climate change affects insured risks as well as demand and supply conditions in insurance markets (Tesselaar et al., 2022).

Most of the financial sector climate risk analyses, including the aforementioned studies, focus on abrupt shocks or natural hazard events (Zou et al., 2023; Reinders et al. (2023). Nevertheless, Reinders et al. (2023) also identify several climate risk stress testing studies that include more gradual climate impacts on overall macroeconomic conditions that are relevant for the financial sector, such as consequences for GDP. Methods used for assessing such macroeconomic effects include Computable General Equilibrium (CGE) models that allow for examining sectoral impacts in economic equilibrium and Integrated Assessment Models (IAMs) of climate change and the economy (Reinders et al., 2023). The Network for Greening the Financial System (NGFS) used IAMs for developing climate risk stress test methodologies, which have been a key input for these type of stress tests produced by financial institutions including central banks, such as the European Central Bank (2021). This methodology involves estimating the physical risks as a proportion of GDP using IAMs with simple damage functions of aggregated economic damages caused by temperature rise. In particular, the NGFS (2021) adopts a country level damage functions that are based on econometric estimates of market impacts of climate change by Kalkuhl & Wenz (2020). Although they offer a relatively simple approach for estimating aggregate climate change losses, the realism of these types of damage functions has been contested (e.g. Botzen et al., 2019; Stern et al., 2022). The main critiques have focused on their aggregate spatial resolution and insufficient consideration of the potential non-linear climate change impacts as well as risks of climate change beyond gradual GDP losses, among other issues (van den Bergh and Botzen, 2015; Fisher-Vanden and Weyant, 2020).

In 2022, the Central Bank of Mexico (BM) started a national assessment of the transition and physical risks climate change could imply for the country's financial sector. For this purpose, the BM selected a suite of three global models that were coupled to produce harmonized sets of future scenarios and risk estimates (Figure 1). The suite of models is composed of two IAMs and a macroeconomic model. The Global Change Analysis Model (GCAM) is a market-equilibrium, integrated assessment model that was used to produce consistent sets of scenarios of variables such as greenhouse gases (GHG) emissions, energy demand and generation, energy production costs and prices, and carbon shadow prices (Calvin et al., 2019; Edmonds & Reilly, 1983). GCAM is developed by the Pacific Northwest National Laboratory (PNNL). CLIMRISK is a spatially explicit, global integrated assessment model developed by the National Autonomous University of Mexico (UNAM) and the VU university Amsterdam (Estrada & Botzen, 2021). As described in detail in the next section, due to the advantages this model offers in comparison with other IAMs, this model was chosen to evaluate the country's physical risks from climate change. Finally, the third model is General Monetary and Multisectoral Macrodynamics for the Ecological Shift (GEMMES), which is developed by the French Development Agency (Bovari et al., 2020; Moreno et al., 2024). It is a stock-flow consistent growth model that, using the GCAM and CLIMRISK output, provides assessments of the macroeconomic impacts transition and physical risk may impose to the Mexican economy. This suite of models produces a comprehensive analysis that encompasses emissions, GDP, transition risks, and macroeconomic factors and provides a robust framework for understanding the multifaceted impacts of climate change on the financial sector.

This paper focuses on describing the assessment of the physical risk that was conducted for Mexico's financial sector. In comparison with similar central banks efforts, Mexico's case study contributes with some useful innovations for the estimation of physical chronic climate risks for the financial system, and a more in-depth analysis of chronic risk. Some of the main innovations is that the analysis is conducted at the grid-cell level with spatially explicit measures of exposure, hazard and sensitivity. This type of analysis allows distinguishing urban and non-urban areas and to estimate the joint effects of local and global climate change in the economy. As has been shown in previous publications, omitting the effects of local climate change in cities due to the urban heat island effect (UHI) leads to severe underestimation of the costs of climate change (Estrada et al., 2017; Estrada & Botzen, 2021). Moreover, the study explores and reports the uncertainty intrinsic to IAMs' damage functions, providing estimates of 18 of them that encompass conservative, high-impact nonlinear, and catastrophic versions of damage functions in the literature. Moreover, the damage function selection includes the three commonly used types of approaches used in the literature for their estimation: enumerative, computable general equilibrium and econometric. Several metrics of damages are used to describe the results, ranging from commonly used to innovative, and covering from a highly granular (grid-cell level), municipality, state, national, regional, and global scales. Economic loss estimates are accompanied by tailor-made, uni- and multivariate risk indices that include diverse climate and economic components.

The remainder of this paper is structured as follows. Section 2 describes the data and methods in the study, including the CLIMRISK model, the modelling decisions, socioeconomic and emissions scenarios, as well as the metrics chosen. The physical risk estimates are reported and discussed in Section 3. Section 4 presents the conclusions and describes some of limitations that we believe need to be prioritized in future studies.

2. Data and methods

*Emissions and socioeconomic scenarios*

The emissions scenarios used in this case study were constructed using the GCAM model (v6) and include 24 GHG and climate pollutant substances that were used to run the climate module in CLIMRISK. The global and regional emissions generated were calculated based on the SSP2 scenario, with a downward adjustment in economic growth for the Mexican economy proposed by the Bank of Mexico. For Mexico's case study four emissions scenarios were used, including the NGFS' Current Policies (CP), Below 2ºC (B2) and Delayed Transition (DT). The fourth scenario was developed to assess the risks of adopting a free rider (i.e., no emissions reduction with respect to the CP scenario) type of national mitigation policy, while the rest of the world follows a trajectory consistent with the Below 2ºC scenario. This scenario is referred to as Asymmetric (FR).

The global GDP and population scenarios used in this study correspond to the SSP2 narrative (Riahi et al., 2017). The spatially explicit (0.5ºx0.5º) GDP and population scenarios used here are those included in CLIMRISK, except for Mexico. In that case, the growth rate of GDP was modified to make the projection more consistent with observed GDP data. The resulting national GDP scenario was made spatially explicit using the dynamic pattern-scaling technique in CLIMRISK (Estrada & Botzen, 2021). The spatially explicit socioeconomic scenarios are available at https://datapincc.unam.mx/datapincc/.

*Brief description of the CLIMRISK model*

CLIMRISK is a spatially explicit, policy evaluation IAM for estimating physical risks of climate change that is structured in four main modules: exposure, hazard, vulnerability and impacts and risk (Figure 1). Each of these modules produces estimates at a spatial resolution of 0.5ºx0.5º worldwide and at an annual frequency, except for the climate module which also produces monthly information. The main objective of CLIMRISK is to offer a simplified and flexible tool that produces tailor-made estimates to characterize different dimensions of the potential consequences of climate change and of the benefits of active climate policy. For this purpose, it produces socio-economic and probabilistic climate scenarios, a variety of metrics of economic damages, as well as tailor-made, dynamic, uni- and multi-variate risk indices. Moreover, it is also integrated with external sectoral impact models such as Xanthos (Dolan et al., 2021; Li et al., 2017), CLIMRISK-River (Ignjacevic et al., 2020), and AIRCCA (Estrada et al., 2020) for producing hydrological, flooding and agricultural scenarios, respectively. The socioeconomic module produces grid-cell level population and GDP scenarios from global, regional or country level time series. This is achieved using dynamic, SSP-dependent, pattern scaling which allows downscaling GDP and population projections to the grid-cell level, maintaining the consistency with aggregated quantities and a SSP narrative (Estrada & Botzen, 2021). The regional/grid-cell downscaling can be produced through three different dynamic scaling patterns that derive from the OEDC-Env group, IIASA and PIK quantifications. The socioeconomic scenarios used in this study correspond to the SSP2 scenario, except for the GDP projection for México which was adjusted downwards by the BM given the slower growth that has been experienced in the past decade. Based on population counts, this module identifies grid-cells that contain existing large urban areas and those that would form in the scenario-dependent future. This module also contains and can integrate RCP land use change scenarios (Riahi et al., 2017) for providing additional context to projected impacts and risks.

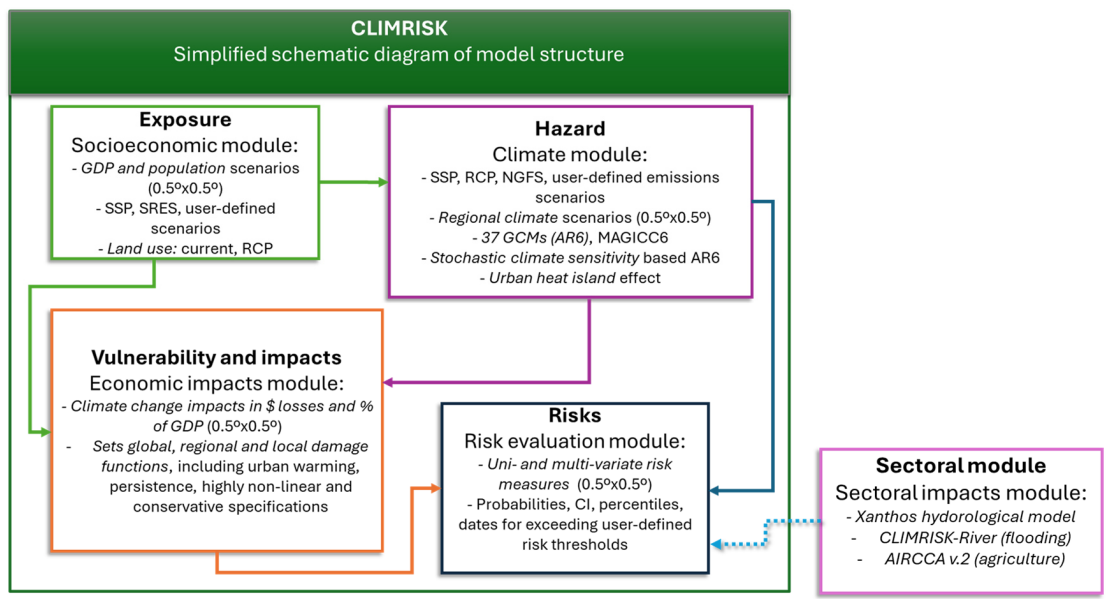

**Figure 1.** Simplified schematic diagram of CLIMRISK's model structure.

The climate module includes a stochastic version of the MAGICC6 reduced complexity climate model. In this version of the model the climate sensitivity parameter is represented by a triangular distribution specified using the likely range and best estimate values from the Intergovernmental Panel on Climate Change Sixth Assessment Report. The emissions scenarios are exogenous to CLIMRISK and in this study come from the GCAM model ran by the BM team. Regional climate projections with a spatial resolution of 0.5°x0.5°are produced by emulating 37 Earth Systems Models included in the CMIP6. The module produces annual and monthly probabilistic estimates of changes in minimum, maximum and mean temperatures, wet-bulb globe temperature and precipitation for each grid cell. In addition to GHG-driven climate change, CLIMRISK generates projections of urban warming due to the UHI effect based on population counts in grid cells that are identified to contain large urban centers. UHI projections include annual and seasonal average and maximum values.

The vulnerability and impacts module incorporates a wide collection of global, regional, and local damage functions, encompassing different estimation approaches: enumerative, computable general equilibrium, and econometric. The base damage functions in CLIMRISK are: 1) the

Kompas (K) damage functions which were derived from recently published estimates of economic losses from climate change (Kompas et al., 2018). The K set includes 178 individual national damage functions and 60,000+ grid-cell level damage functions; 2) the Kalkhul and Wenz (KW) panel and cross-sectional damage functions (Kalkuhl & Wenz, 2020); 3) the RICE2010 (R) regional damage functions (Nordhaus, 2010) and the DICE2016R (d) damage function (Nordhaus, 2017), and; 4) the Weitzman (w) catastrophic global damage function (Weitzman, 2009). In addition to these original damage functions, CLIMRISK includes extensions to some of them to include the effects of persistence (Estrada et al., 2015) of impacts and the UHI (Estrada et al., 2017). Moreover, CLIMRISK extends regional damage functions to the grid cell level in a way that the aggregated damage is consistent with the original projected losses at the regional level (see SI in Estrada & Botzen, 2021 for a detailed description). Since the previous version of CLIMRISK, which was limited to the R-based damage functions with added persistence and/or UHI effect, the vulnerability and impacts module was able to cover most of the uncertainty in impact projections reported in the literature (Estrada & Botzen, 2021; Howard & Sterner, 2017).

The current version of CLIMRISK offers a wider range of damage functions, as well as the newly available option of downscaling global and regional damage functions to grid-cell level, besides enriching global and regional projections with local damage functions (upscaling). CLIMRISK allows downscaling such projections at the regional and grid-cell using the spatial patterns created by a user-selected set of regional damage functions. Note that these spatial patterns are time-dependent, as well as SSP/RCP path-dependent, making them unique for each simulation. With this approach, damage functions for catastrophic damages from climate change which generally have no regional dimension can be downscaled at the regional/local scales in CLIMRISK. In the current version of CLIMRISK, the grid-cell version of RICE2010 damage functions is used to downscale the damages produced by the catastrophic global damage function proposed by Weitzman (Weitzman, 2009). Similarly, this approach can also be used to update regional/local projections to more recent estimates of global damages. For instance, latest version of the Nordhaus models is the DICE2016R, and its output is used to scale the projections obtained with the RICE2010 regional damage functions, updating the regional RICE2010 projections to be consistent at the global scale with those of DICE2016R. Table 1 provides a list of the damage functions used for the estimates presented in this paper and Figure 2 illustrates the range of projected damages for a warming of up to 2.9ºC in global temperatures.

Several metrics are included in CLIMRISK to gain further insights from the damage estimates produced by CLIMRISK, and to facilitate the communication of results to users and decision-makers. Economic loss estimates in CLIMRISK have an annual frequency and are produced at the grid-cell level and can be aggregated at the subnational (state and municipality), national, regional, and global levels. The metrics that are used here are: the yearly losses as a percentage of the corresponding GDP value; the present value of losses over the period 2010-2100; the relative present value which is the ratio of the present value estimate and a reference GDP value, commonly that of the current year (i.e., 2024); rolling window present value calculated for a window size that represents a cycle of interest, such as governmental office terms and the length of investment decisions; relative risk change estimates that are calculated as the ratio between the first rolling window estimate and those at future periods, aiming to provide how physical risk would evolve for specific lengths of investment decisions and time horizons.

**Table 1.** List of damage functions and extensions used for estimating economic losses for this study.

| Damage function | CLIMRISK's extensions used in this study | | | |
|---|---|---|---|---|
| *Name and ID* | *UHI effects* | *Persistence* | *Update* | *Downscaling* |
| Kompas: K | KU | -- | -- | -- |
| RICE2010: R | RU | RPU | $R_d$, $RU_d$, $RPU_d$ | -- |
| Kalkuhl and Wenz: KW (panel) | KWU | -- | -- | -- |
| Weitzman: w | $RU_w$ | $RPU_w$ | -- | R |

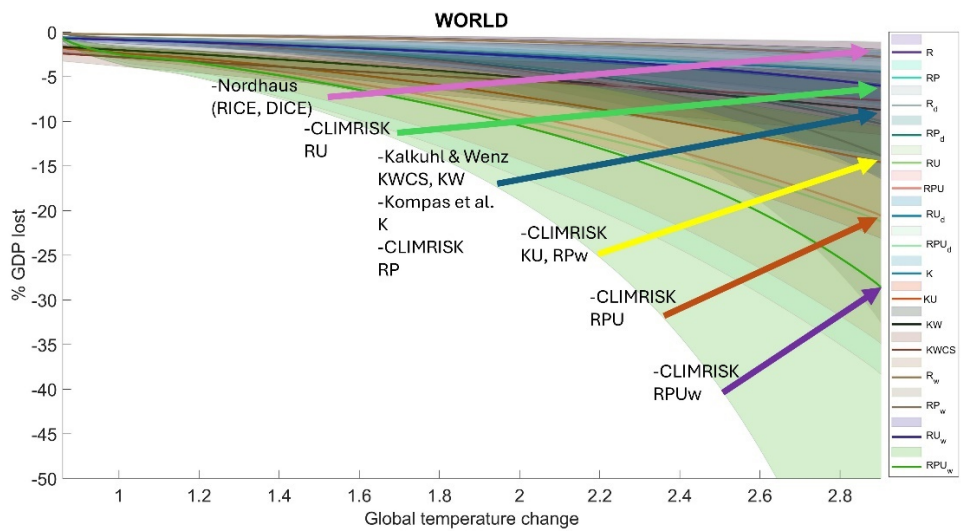

**Figure 2.** Illustrative representation of the projected global losses as a percentage of GDP for a selection of damage functions in CLIMRISK and for the CP and SSP2 scenarios. Colored lines represent the estimated median loss, while the respective shades show the 90% confidence intervals.

The risk module integrates the information generated by the other modules in CLIMRISK, as well as of external sectoral modules, to provide the user a more complete picture of the physical risks climate change that is not restricted to only economic losses. This module produces uni- and multivariate risk indices that combine projections of climatic, economic losses, land use, population and GDP, crop yield changes, water availability and flooding. The risk indices are user-defined, and tailor made for addressing particular information needs. The risk measure include marginal/joint probabilities of exceedance and dates for exceedance of user-defined risk thresholds, confidence intervals, identification of hotspots, time of emergence of economic impacts (Ignjacevic et al., 2021), and percentile estimates.

## 3. Results and discussion

*Climate change projections for the selected emissions scenarios*

The results presented in this section are based on simulation experiments of 500 realizations for each of the emissions scenarios that were selected. The projected global mean annual temperature changes for the periods 2040-2060 and 2080-2100 are reported in Table 2. The largest changes in global temperature occur under the CP scenario with a mean increase of 2.70°C during the 2080-2100 period, with respect to preindustrial values. When considering the 95$^{th}$ percentile the CP could reach 3.45°C warming by the end of the century. The CP scenario projects that the goal of limiting warming below 2°C would be exceeded by 2050. As expected, the B2 and FR lead to the same global temperature change projection when rounding to two digits, as the difference between the two is that Mexico does not adopt further emissions reductions, but the rest of the world follows the B2 path. The DT trajectory implies slightly larger warming than B2 and FR, but the 90% confidence intervals show an almost complete overlap. While the B2, FR and DT provide interesting insights for transition risks, they all convey very similar information for physical risk assessment. As such, for the remainder of this paper the results and discussion will focus mostly on the CP and B2 scenarios, and the results of the FR and DT scenarios are available upon request.

**Table 2.** Global mean annual temperature change for the selected emission scenarios

| Emissions scenario | 2040-2060 | 2080-2100 |
|---|---|---|
| CP | 2.34 (1.38, 2.34) | 2.70 (1.91, 3.45) |
| B2 | 1.72 (1.24, 2.13) | 1.75 (1.18, 2.27) |
| DT | 1.77 (1.29, 2.21) | 1.79 (1.22, 2.34) |
| FR | 1.72 (1.24, 2.13) | 1.75 (1.18, 2.27) |

Figures are expressed in °C. 5$^{th}$ and 95$^{th}$ percentiles are reported in parenthesis.

The CP scenario implies high probabilities of exceeding significant thresholds that could endanger human and natural systems. Figure 3a shows the dates for exceeding a warming of 3°C per grid cell in the world, and the estimated dates for Mexico range from 2080 in the northern part of the country to 2090-2100 for the central and southern regions. This threshold was chosen as the literature suggests that regional changes of similar magnitude and velocity of change in mean annual temperature would put in significant danger various ecosystems (Estrada & Botzen, 2021; Fischlin et al., 2007). The literature has also shown that decreases in precipitation and drought under warmer climates are linked to increasing social problems such as migration and conflict (Hodler & Raschky, 2014; Hsiang et al., 2013; Puente et al., 2016; World Bank, 2016), as well as declining ecosystems services, wildfires and negative effects on biomass production (De Dato et al., 2008; McDowell & Allen, 2015; Piñol et al., 1998; Stevens-Rumann et al., 2018). Figures 3b-3c show the dates for exceeding a reduction of at least 10% in annual precipitation, and the dates for jointly exceeding both 3ºC in warming and at least a 10% decrease in annual precipitation, respectively. By mid-century, the northwestern part of Mexico would exceed both risk thresholds and for the southern part of the country, including the Yucatán peninsula, the joint exceedance of these thresholds could happen about 15 years earlier. Both regions contain ecosystems of importance for biodiversity conservation and the existence of threatened endemic species. The risks of exceeding these marginal and joint thresholds for Mexico can be avoided during this century under the B2 scenario. This is also the case for any other scenario that is consistent with the Paris Agreement goals, such as the FR and DT.

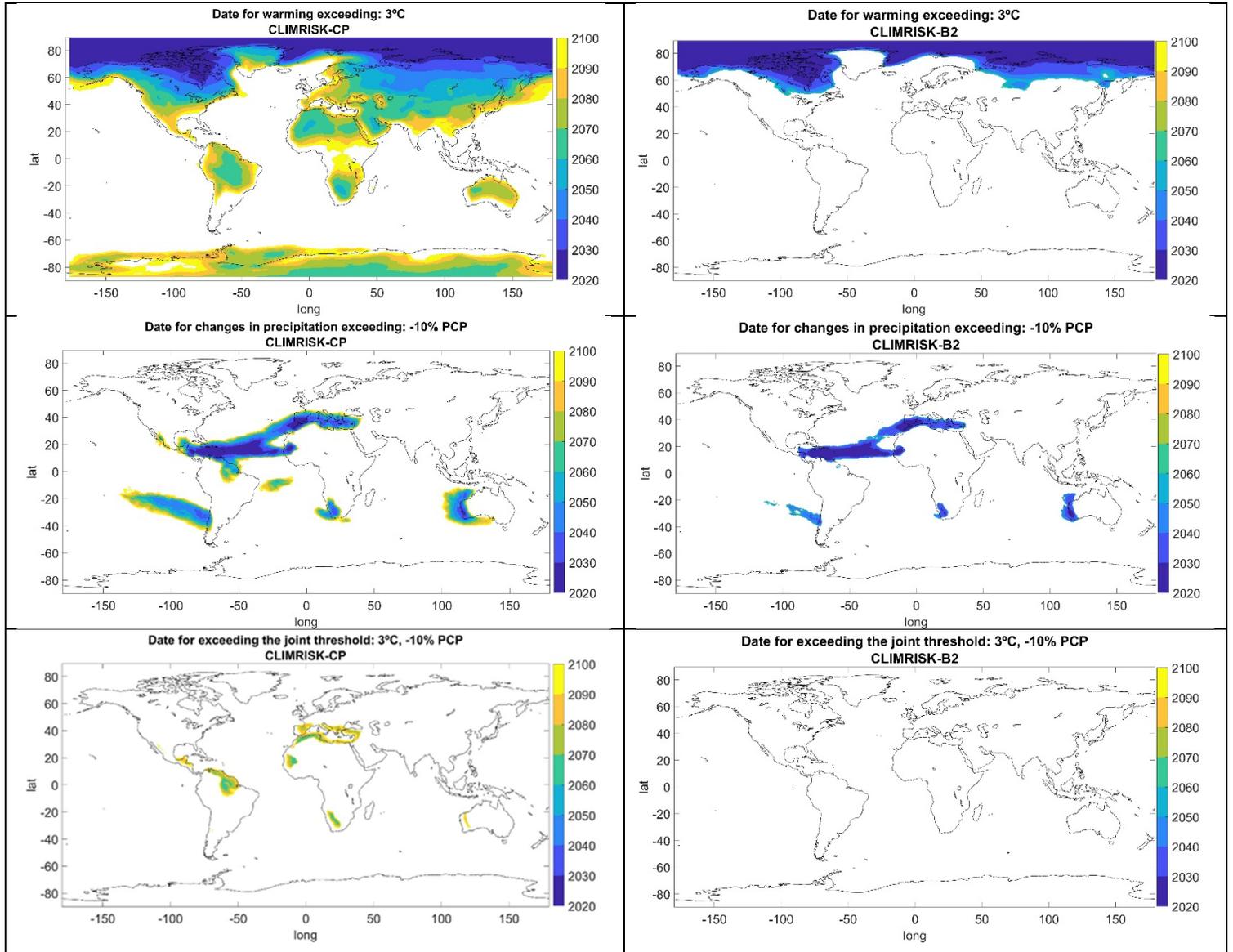

**Figure 2.** Dates for exceeding selected risk thresholds in levels of warming and precipitation change. The upper row shows the dates for exceeding 3°C increase in annual temperature, while the middle row shows the dates for exceeding a decrease of at least 10% in annual precipitation. The lower row shows the date for jointly exceeding both warming and precipitation thresholds. The left and right columns show the results for the CP and B2 scenarios.

*Economic damages under the CP and B2 scenarios*

As shown in Figure 3, the economic losses from climate change in the CP scenario could represent more than 35% of Mexico's GDP in 2100, considering the estimated 90% confidence intervals. The Nordhaus' damage functions (DICE/RICE) are the most conservative and project less than 2% loss of GDP at the end of the century, while the RPUw is highly nonlinear on temperature increase and leads to the largest losses.

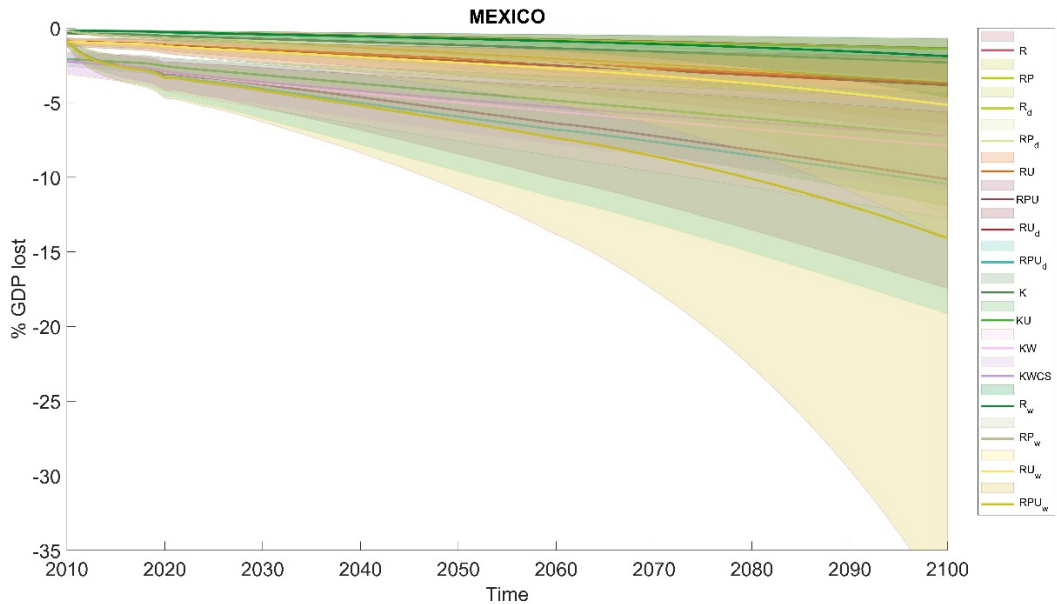

**Figure 3.** Projected losses for Mexico as a percentage of its GDP for a selection of damage functions in CLIMRISK and for the CP and SSP2 scenarios. Colored lines represent the estimated median loss, while the respective shades show the 90% confidence intervals.

Table 2 shows the present value estimates of damages over this century for a selection of damage functions and a 1.5% discount rate. The selection of damage functions ensures that much of the range of estimates that are available in the literature are represented. It only discards the original R damage functions, which are known to omit important aspects such as UHI and persistence and that are based on possibly outdated information. Note that the reported uncertainty ranges are large and can be uncomfortable for decision making. However, this case study favors presenting all the estimates deemed credible by our team of experts to the involved stakeholders in order not to withhold any information that could be relevant to them. Unjustified reductions of uncertainty can lead to trading information for ignorance and biasing decisions (Gay & Estrada, 2010). Considering the central estimates of the different damage function, the present value of losses for Mexico under the CP scenario ranges from $2.1 (K) to $19.1 (KWU) trillion dollars, while if the 90% confidence intervals are included then the present value estimates are in the range of $1.3-$27.5 trillion. Mexico's losses are quite significant, accounting for about 20% of the sum of those of the rest of Latin America, and less than 1% of the total global losses. This study also found that communication of results can be improved if the estimates are reported in units which users of the information are familiar with. For instance, reporting the present value estimates of losses as units of the country's GDP helps understand their magnitude. As is shown by the numbers in brackets in Table 2, the economic losses from climate change during this century would represent today between 1.1 to 10.7 times Mexico's GDP in 2024 (or between 71% and 15.4 times its 2024 GDP, considering the 90% confidence intervals).

Under the B2 scenario, which results from an international mitigation effort consistent with the Paris Agreement, the losses for Mexico can be significantly reduced. The implementation of the B2 scenario would represent benefits for Mexico, in terms of avoided losses, between 0.38 and 2.5 times the current GDP. Considering all damage functions in Table 2, the range of confidence intervals is asymmetric ranging from 0.23 to 7 times the current GDP in avoided losses. However, even under such ambitious international mitigation effort, the resulting residual impacts are high. The impacts under the B2 scenario for Mexico are in the range of 81% to 9 times GDP in 2024, with the confidence intervals extending from 50% to 12.4 times 2024's GDP.

**Table 2.** Total discounted economic costs of climate change over this century expressed in billions US$2005 and as a percentage of Mexico's GDP in 2024, under the CP and SSP2 scenarios for a selection of damage functions in CLIMRISK.

|    | K | RUd | RUw | KU | KW | RPUd | RPUw | KWU |
|----|---|-----|-----|-----|-----|------|------|-----|
| CP | 2117 [1.19] (1274, 3034) [0.71, 1.7] | 3873 [2.17] (2150, 6571) [1.21, 3.68] | 4443 [2.49] (2228, 9362) [1.25, 5.25] | 7547 [4.23] (4965, 11459) [2.78, 6.43] | 8561 [4.8] (5434, 13210) [3.05, 7.41] | 10395 [5.83] (5788, 17585) [3.25, 9.86] | 11941 [6.7] (5999, 25140) [3.36, 14.1] | 19088 [10.7] (13414, 27515) [7.52, 15.43] |
| B2 | 1451 [0.81] (856, 2000) [0.48, 1.12] | 2687 [1.51] (1489, 4320) [0.84, 2.42] | 2786 [1.56] (1511, 4691) [0.85, 2.63] | 5827 [3.27] (3916, 8376) [2.2, 4.7] | 6807 [3.82] (4228, 10248) [2.37, 5.75] | 7285 [4.08] (4042, 11694) [2.27, 6.56] | 7556 [4.24] (4102, 12710) [2.3, 7.13] | 15888 [8.91] (11205, 22127) [6.28, 12.41] |
| AL | 666 [0.38] (418, 1034) [0.23, 0.58] | 1186 [0.66] (661, 2251) [0.37, 1.26] | 1657 [0.93] (717, 4671) [0.4, 2.62] | 1720 [0.96] (1049, 3083) [0.58, 1.73] | 1754 [0.98] (1206, 2962) [0.68, 1.66] | 3110 [1.75] (1746, 5891) [0.98, 3.3] | 4385 [2.46] (1897, 12430) [1.06, 6.97] | 3200 [1.79] (2209, 5388) [1.24, 3.02] |

Note: all numbers refer to estimates in US$2005, except for those in brackets which are expressed in units of Mexico's GDP in 2024. 90% confidence intervals are presented in parenthesis and brackets for US$2005 and units of GDP in 2024, respectively. Discount rate: 1.5%.

The explicit spatial resolution of CLIMRISK makes it possible to project damages in a global 0.5ºx0.5º grid. This is possible because the model projects exposure, hazard and includes damage functions that are specified at the grid cell level. The level of granularity offered is useful, for example, to identify geographical areas in which investments are located and physical risk is high. As it is discussed in the next sections, the level of granularity and the variety physical risk related variables in CLIMRISK allows for identifying multivariate, climate-economy risk hotspots.

Figure 4 shows the relative present value of losses for Mexico at the grid-cell level using the KWU and KU damage functions. This figure illustrates the large spatial heterogeneity that characterizes the economic impact of climate change, and that the uncertainty is not only related to the magnitude of losses but also how they are distributed among and within countries. Both the KWU and KU project that areas with large urbanizations in Mexico would experiment the largest losses relative to their GDP, although the projected magnitudes vary greatly between them. These areas with high physical risk correspond to the megalopolis of the Valley of Mexico, which includes CDMX, Puebla, Tlaxcala, Estado de México and Hidalgo, the industrial corridor of Querétaro, Jalisco, Michoacan, Aguascalientes, San Luis Potosí, as well as the Nuevo León and Coahuila. Nonetheless, KWU projects more generalized areas at high risk along the coast of the Gulf of Mexico and for the Yucatán peninsula. It is important to note that omitting the UHI effect not only biases downwards the aggregated losses form climate change as has been discussed previously, but those biases also distort the geographical distribution of losses, severely underestimating the risks for urban areas (Figure S1).

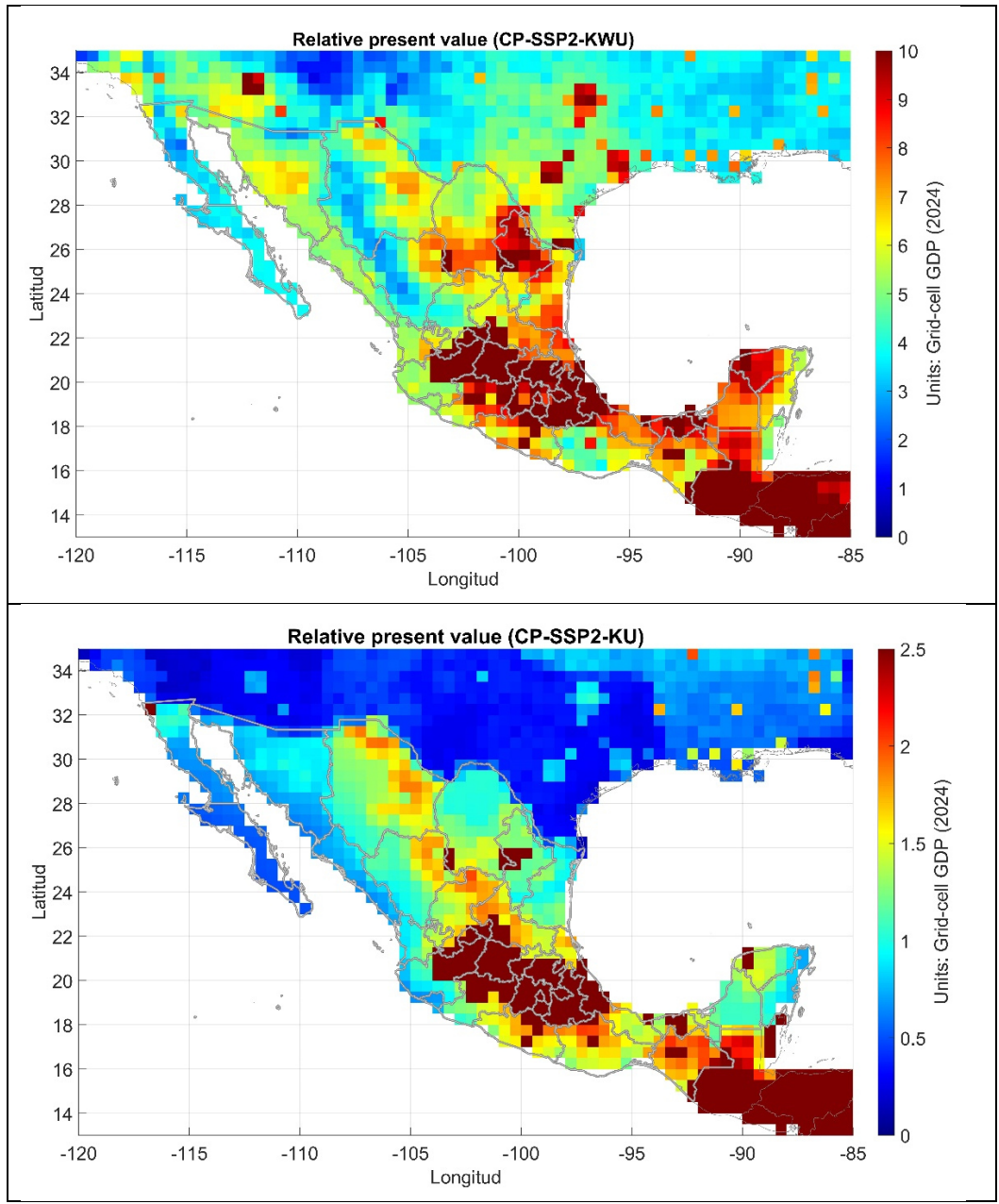

**Figure 4.** Spatial distribution of economic losses from climate change. Panel a) shows the distribution of the relative present value of economic losses during this century using the KWU damage function, while panel b) shows the relative present value of losses using the KU damage function. Discount rate: 1.5%.

*Improving information use: metrics for the short term, risk indices and hotspots, and investment-risk mapping at the localities level*

Introducing the chronic physical risks associated with climate change into the financial system's evaluations has been shown to be challenging. Part of this resides in the differences between temporal horizons for evaluating climate change's risk and those typically used in the financial system. In this section we introduce a novel metric based on the ratio of rolling window present value estimates with respect to a reference year. These rations of rolling window estimates allow inferring the systematic risk climate change could introduce to commonly used investment horizons though the impacts of the phenomenon on the economy.

Figure 5 illustrates for an investment horizon of 5 years how the systematic climate risk would increase according to the KWU and KU damage functions under the CP-SSP2 scenario. Taking 2024 as the initial estimate for a five-year horizon investment, carrying out the same five-year period investment one year later would involve an additional 5% increase in systematic risk climate change for the economy. The same five-year investment in 2030 would carry on an additional 28%-30% increase in climate risk for the economy, in comparison with 2024. By mid-century, the additional climate risk could reach 138%-161%. An important finding is that, even if the B2 emissions scenario is implemented at the global level, the additional climate related systematic risk for a five-year investment horizon would be very similar to that of the CP scenario, suggesting that supplementary risk reduction would be necessary.

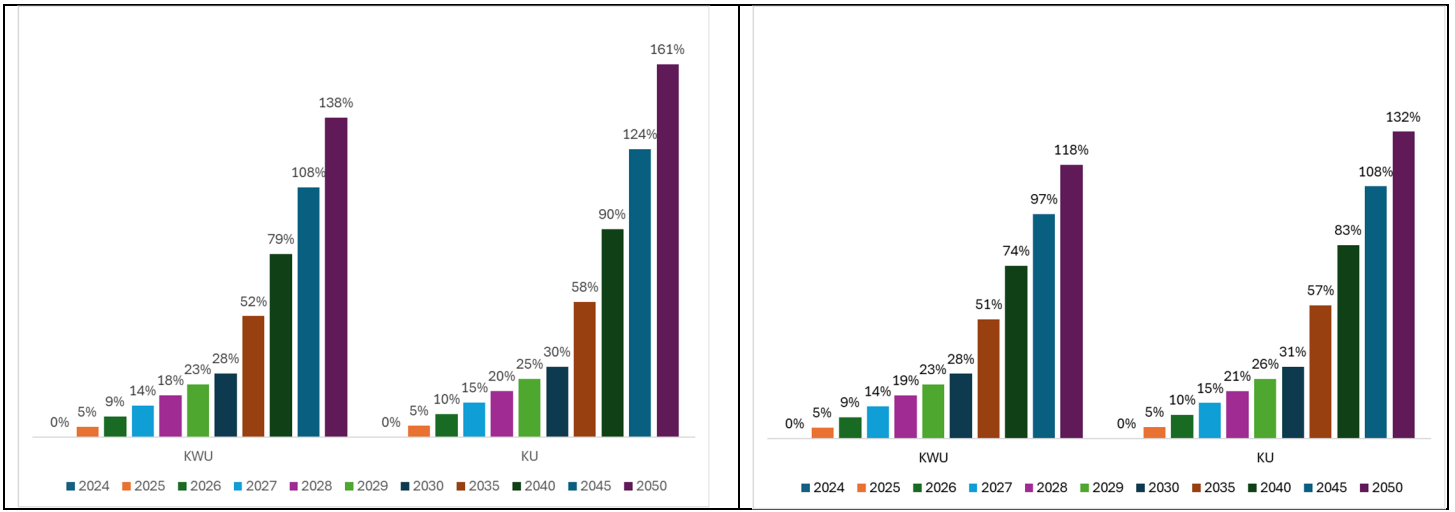

**Figure 5.** Relative systematic risk increases from climate change for five-year rolling window horizons. The left panel shows the change in additional systematic risk for the CP scenario, while the right panel shows such estimates for the B2 scenario.

CLIMRISK produces uni- and multivariate risk indices which combine climate and economic projections, and user-defined risk thresholds for constructing a more comprehensive view of the consequences of climate change. These indices can help users to personalize results according to their information needs and to visualize geographical areas in which different risks converge. Figure 6 provides an illustration of a multivariate risk index that is based on the following user selected thresholds applied at the grid cell level: exceeding 3ºC in warming; decrease of at least 10% in annual precipitation; exceeding losses of at least 10% per year; exceeding annual losses of more than 1 billion dollars. The upper panels of Figure 6 show the dates for reaching moderate risk levels (jointly exceeding at least two of these thresholds), and the lower panels presents the dates reaching high risk levels (jointly exceeding at least three of the selected thresholds). The left column presents the results for the CP scenario and shows that for places located in Africa, Asia and India, the levels of moderate risk were attained during the past decades. In the case of Mexico, the central part of the country and the northern coast of the Pacific Ocean will exceed such level during this decade, while the rest of the territory will do so during the last thirty years of this century. The levels of high risk (lower-left panel) would be attained in the central part of Mexico before 2050 and near the end of this century for a considerable fraction of the country. The left column shows the dates for reaching moderate and high levels of risk under the B2 scenario. At the global scale, there is a widespread decrease in the total area of the world that would exceed the moderate- and high-risk thresholds, except for regions of the central part of Mexico, as well as the Arctic, areas in China, India, Africa, and the Mediterranean, among others. This analysis allows to identify that the main urban areas in Mexico have already exceed high levels of risk in this multivariate index, and also helps to identify the location of current and future risk hotspots.

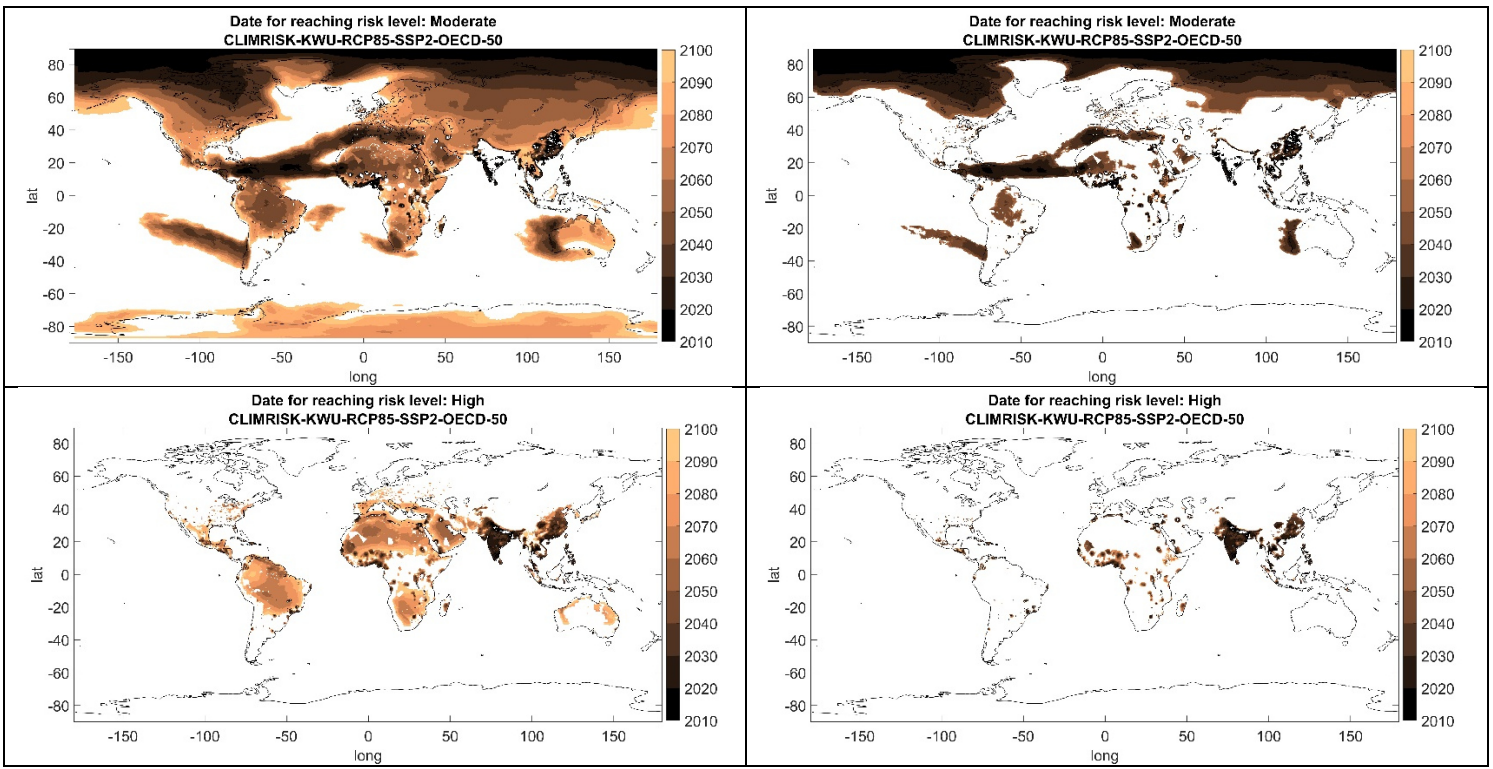

**Figure 6.** Dates for reaching moderate and high risk levels under the CP and B2 scenarios. The upper panels show the results for moderate risk levels for the CP (left) and B2 (right) emissions

scenarios. The lower panels show the results for high risk levels for the CP (left) and B2 (right) emissions scenarios.

The spatial granularity at which CLIMRISK resolves exposure, hazard and sensitivity, as well as its ability to provide additional information regarding agriculture, hydrology and land use can help decision-makers gain a more complete description of the localities where they have investments. For the case study of Mexico, a large database of output from CLIMRISK and external modules was compiled to help stakeholders in the Mexican financial system better understand the level of physical risks climate change entails for the country. An interactive interface was designed to map these data base in localities where the financial system currently has investments. As such, users can choose the relevant variables and thresholds to construct risk indices for their particular needs and find the level of risk associated to the localities of interest in which the financial institution has investments.

## 4. Conclusions

The creation of frameworks and models for helping financial institutions integrate physical risks into their risk assessments and management is an active area of research. IAMs have been identified as relevant tools for this purpose. However, some of the shortcomings of most of the available IAM, have made difficult their use. Among such limitations are their lack of spatial granularity, too simplistic representations of uncertainty, their dependence on a very limited number of damage functions, differences in time horizons between climate and financial analysis, as well as the scarcity of methods to connect chronic physical risk to financial risk assessments.

The case study conducted by the Bank of Mexico includes some innovations in its framework design and in the integration and adaptation of tools that help addressing some of the common limitations in this type of efforts. One salient strength of the analysis consists in the integration of three specialized models which allow for a consistent estimation of the transition and physical risks, and of a detailed analysis of the macroeconomic consequences of such risks. With respect to physical risks, the study uses a new generation IAM which has unique features such as being spatially explicit, to emulate 37 Earth System Models, to incorporate the local warming and impacts caused by the urban heat island effect, and to simulate damages for a wide set of damage functions that are encompassing of the literature. Moreover, the Bank of Mexico promoted a close collaboration between financial experts and the selected modelling groups to fine tune the modeling outputs to better address the information requirements of the users in the Mexican financial system.

The assessment of the physical risks for Mexico reported in this study shows that the consequences of unabated climate change can severely affect the economy and introduce significant additional stress the financial system. The analysis, datasets and tools generated for this study provide an important step forward in better understanding the risks and consequences of climate change for the Mexican financial system, as well as a steppingstone for future research in the country and other regions.